\documentclass[twocolumn,aps,a4]{revtex4}
\usepackage{graphicx}

\begin{document}
\date{\today}
\title{Getting the elastic scattering length by observing inelastic collisions in ultracold metastable helium atoms}
\author{S. Seidelin, J. Viana Gomes\footnote{Permanent address: Departamento de Fisica, Universidade
do Minho, Campus de Gualtar, 4710-057 Braga, Portugal}, R.
Hoppeler, O. Sirjean, D. Boiron, A. Aspect and C. I. Westbrook}
\affiliation{\mbox{Laboratoire Charles Fabry de l'Institut
d'Optique, UMR 8501 du CNRS, F-91403 Orsay Cedex, France}}

\begin{abstract}

We report an experiment measuring simultaneously the temperature
and the flux of ions produced by a cloud of triplet metastable
helium atoms at the Bose-Einstein critical temperature. The onset
of condensation is revealed by a sharp increase of the ion flux
during evaporative cooling. Combining our measurements with
previous measurements of ionization in a pure BEC,
we extract an improved value of the scattering length
$a=11.3^{+2.5}_{-1.0} $~nm. The analysis includes corrections
taking into account the effect of atomic interactions on the
critical temperature, and thus an independent measurement of the
scattering length would allow a new test of these calculations.
\end{abstract}
\pacs{03.75.Hh, 34.50.-s, 67.65.+z}\maketitle

Understanding and testing the role of interparticle interactions
in dilute Bose-Einstein condensates (BEC) is an exciting area of
current research. Although measurements of the interaction energy
and the spectrum of excitations of a BEC have confirmed the
validity of the Gross-Pitaevskii equation \cite{rev_stringari},
there are still relatively few quantitative tests of other
aspects, such as the effect of interactions on the value of the
critical temperature ($T_{\rm c}$) or the condensed fraction
\cite{fabrice,cornell}. The success in condensing metastable
helium atoms (He*) \cite{BecHe,BecENS}, was greeted with interest
in the community partly because the metastability offers new
detection strategies unavailable with other species. To fully use
these strategies
however, we are still missing an accurate value of the s-wave
scattering length $a$, the atomic parameter which determines all
elastic scattering behavior at low energies. An accurate value of
$a$ would also be useful to help clarify some puzzling results
concerning measurements of He* in the hydrodynamic regime, in
which two different ways of measuring the elastic scattering rate
appeared to be in contradiction \cite{acta}. Also, because He is a
relatively simple atom, theoretical predictions of $a$ are already
in a rather narrow range \cite{Gadea,Leo:2002} and these
calculations should be tested.

A straightforward method to determine $a$ is to use ballistic
expansion of a BEC to measure the chemical potential for a known
atom number. This was done in Refs.~\cite{BecHe,BecENS}, but the
measurements were limited by the calibration of the number of
atoms, or equivalently the density of the sample. The reported
values for $a$ are $20\pm 10$~nm and $16\pm 8$~nm, respectively. A
more recent estimate, limited by similar effects, is $a=10\pm
5$~nm \cite{Tol}. In this paper we report a new measurement of $a$
which makes extensive use of a unique feature of He*, spontaneous
Penning ionization within the sample.

We exploit two specific situations in which the absolute atom
number $N$ is simply related to $a$ and measured quantities:
({\it{i}}) for a pure BEC, the number is deduced directly from the
chemical potential $\mu$ and $a$, ({\it{ii}}) for a cloud at the
Bose-Einstein threshold it is simply related to the critical
temperature $T_{\rm c}$. Both $\mu$ and $T_{\rm c}$ are accurately
deduced from time of flight (TOF) measurements. Comparison of ion
rates from a pure BEC of known chemical potential and from a cloud
at $T_{\rm c}$ allows us to extract $a$ and the ionization rate
constants. The deduced value of $a$ is independent of the absolute
ion detection efficiency, assuming that this (poorly known) efficiency
is the same in the two measurements. The ion signal is also used
in another novel way: Since it provides a real-time observation of
the onset of BEC \cite{Seidelin}, we use it to reliably produce a
cloud at the condensation threshold.

A dense cloud of He* produces a steady flux of ions due to various
ionization processes. Density losses due uniquely to
{\it{ionizing}} collisions depend on the local density $n$
according to: $ \left(\frac{dn}{dt}\right)_{ionizing} = -
\frac{n}{\tau_{i}} - \beta \, n^{2} - L \, n^{3} $ with $\tau_{i}$
the lifetime due to ionizing collisions with the background gas
and $\beta$ and $L$ the 2-body and 3-body ionization rate
constants defined for a thermal cloud. The total ion rate from a
thermal cloud is given by:
\begin{equation}\label{formulegene}
\Phi = \frac{N}{\tau_{i}} + \frac{1}{2} \, \beta \int n^2
d{\textbf{r}} + \frac{1}{3} \, L \int n^3 d{\textbf{r}}
\end{equation}
The numerical factors reflect the fact that although 2 (3) atoms
are lost in 2-body (3-body) collisions, only 1 ion is produced.
Ionization measurements on a pure BEC were reported in
\cite{PRLHe}, and, as $a$ was not precisely known, $\beta$ and $L$
were given in terms of $a$.

For a precise measurement of $a$, corrections due to interactions
must be taken into account. In the mean field approach, the
density is given by \cite{rev_stringari}:
\begin{equation}\label{formuledensite}
n({\textbf{r}}) = \frac{1}{\lambda^3(T)} g_{3/2}
\left[\exp\left(-\frac{1}{k_B T} \, (V({\textbf{r}})+2 g \,
n({\textbf{r}}) - \mu)\right)\right]
\end{equation}
with $\lambda(T)$ the thermal de Broglie wavelength, $T$ the
temperature of the cloud, $k_B$ the Boltzmann constant, $V$ the
trapping potential energy, $g=4\pi\hbar^2a/m$ the interaction
constant, $\mu$ the chemical potential and $g_{\alpha}(x)
=\sum\limits_{i=1}^{\infty} \frac{x^i}{i^{\alpha}}$.

The ion rate at the phase transition $\Phi_{\rm c}$ can be derived
from Eq.~(\ref{formuledensite}) by a first order perturbation
theory similar to Ref.~\cite{Stringari} but with a fixed
temperature rather than a fixed atom number. We use the chemical
potential of a gas in a harmonic potential at the BEC transition:
\begin{equation}\label{mu}
\mu_{\rm c}/k_{B}T_{\rm
c}=\frac{3}{2}\frac{\widetilde{\omega}}{\omega_{\rm c}
}+4g_{3/2}(1)\frac{a}{\lambda_{\rm c}}
\end{equation}
This gives:
\begin{equation}\label{formuleflux}
\begin{array}{ll}
\Phi_{\rm c} = (\frac{\omega_{\rm c}}{\overline{\omega}})^3 \times
&\left[ \right.  \frac{1}{\tau_{i}} (1.20 + 2.48
\frac{\widetilde{\omega}}{\omega_{\rm c} } + 12.35
\frac{a}{\lambda_{\rm c}})
\\
 &  + \frac{\beta}{\lambda_{\rm c}^3} (0.33 + 1.81
\frac{\widetilde{\omega}}{\omega_{\rm c} } + 6.75
\frac{a}{\lambda_{\rm c}})
\\
 & \left. +\frac{L}{\lambda_{\rm c}^6} (0.22 + 2.21
\frac{\widetilde{\omega}}{\omega_{\rm c} } + 6.50
\frac{a}{\lambda_{\rm c}}) \right]
\end{array}
\end{equation}
with $\widetilde{\omega} =
(2\omega_{\perp}+\omega_{\parallel})/3$,
$\overline{\omega}=(\omega_{\parallel} \omega_{\perp}^{2})^{1/3}$,
$\omega_{\rm c} = k_BT_{\rm c}/\hbar$ and $\lambda_{\rm c} =
\lambda(T_{\rm c})$. The numerical values come from the
calculation of arithmetic series and are independent of any
parameters of the cloud.
The terms proportional to $a/\lambda_{\rm c}$ in
Eq.~(\ref{formuleflux}) account for the atomic interactions, while
the corrections proportional to $\widetilde{\omega}/\omega_{\rm
c}$ take into account the effect of finite sample size.
For the typical parameters of our experiment we have ($T_{\rm c}
\sim 2~\mu$K and $a=12$~nm) $a/\lambda_{\rm c}\simeq
\widetilde{\omega}/\omega_{\rm c} \simeq 0.02$ corresponding to an
interaction correction of $20\%$, $40\%$ and $60\%$ in the three
successive terms in Eq.~(\ref{formuleflux}). Even though the first
order corrections are large, we find, using an approach similar to
Ref.~\cite{Arnold2}, that the second order corrections are
negligible: $-4\%$, $1.8\%$ and $-3\%$, respectively, with the
above parameters. In our case, finite size corrections are always
smaller than those due to interactions; since $a/\lambda_{\rm
c}\simeq \widetilde{\omega}/\omega_{\rm c}$, the difference
corresponds simply to the numerical constants in
Eq.~(\ref{formuleflux}).

Our setup has been described in Ref.~\cite{PRLHe}. Briefly, we
trap up to $2\times 10^8$ atoms at 1~mK in a Ioffe-Pritchard trap
with a lifetime ($\tau$) of 70~s, and a lifetime due to ionizing
collisions, ($\tau_{i}$), estimated to be $>500$~s. In a typical
run, forced evaporation for 30~s cools a cloud to a temperature
near the phase transition. At this point, the rf-knife frequency
is decreasing at a rate of 400 kHz/s. We use a cloverleaf
configuration with a bias field $B_0 =300$~mG. The axial and
radial oscillation frequencies in the harmonic trapping potential
are $\omega_{\parallel}/2\pi=47\pm3$~Hz and
$\omega_{\bot}/2\pi=1225\pm20$~Hz respectively. A 2-stage, single
anode microchannel plate detector (MCP) is placed 5~cm below the
trapping region. Two grids above the MCP allow us either to repel
positive ions and detect only the He* atoms, or to attract and
detect positive ions produced in the trapped cloud. As explained
in Ref. \cite{PRLHe}, to detect the ion flux, the MCP is used in
counting mode, whereas we record the TOF signal at low gain
(analog mode) to avoid saturation. As explained in Ref.
\cite{BecHe}, the TOF signal is due to atoms in the $m=0$ state
which are insensitive to the magnetic field. However, atoms in
magnetic field sensitive states are still present, and their
trajectories are affected by uncontrolled residual fields.
Therefore, during the time of flight, we apply a magnetic gradient
in order to push these atoms away from the detector. The ratio
between the detected atoms in the $m=0$ state and the initial
number of atoms in the cloud is not well known \cite{Seidelin}, so
we use the TOF only to get the temperature.

\begin{figure}
\begin{center}
\includegraphics[height=5cm]{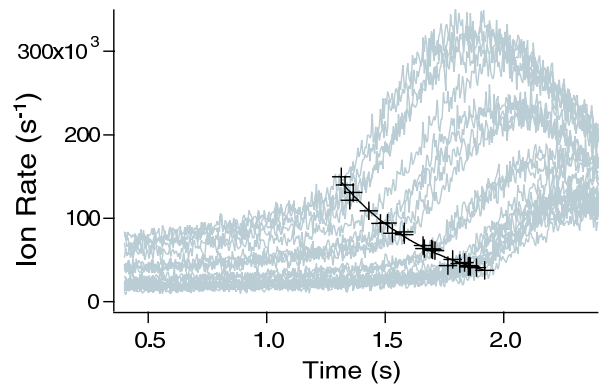}
\end{center}
\caption{Variation of the ion rate as the atomic cloud is cooled
through the phase transition for various initial densities (gray
curves). The rf-knife frequency at $t=0$ is 2~MHz.
The sudden increase of the ion rate (crosses) occurs at the BEC
transition. The solid line passing through the transition points
constitutes our empirical relation, named threshold curve.}
\label{Ionrates}
\end{figure}

The crux of the experiment is to obtain a cloud of atoms at the
phase transition. To identify the BEC threshold point, we monitor
the ion signal. We have shown in Ref. \cite{Seidelin} that the onset
of BEC is heralded by a sudden increase of the ionization rate
associated with the increased density of the condensate. More
precisely, the BEC threshold corresponds to the rapid change in
slope of the ion rate vs time, or the maximum of the 2nd
derivative \cite{th_olivier}. Figure \ref{Ionrates} shows a series
of such ionization rates during evaporation through the BEC
transition. From these curves we can determine an empirical
relation between the time of the onset of condensation and the ion
rate preceding it. This relation stays valid only as long as we
keep the same evaporation ramp and bias field. We will refer to
this as the "threshold curve" in the following. Due to
fluctuations of the bias field, we observe fluctuations of the
time of BEC onset from run to run. These correspond to
approximately $\pm$~60~ms in time or $\pm$~25~kHz in frequency, a
value which agrees with independent measurements of the
fluctuations of the bias field.

Having established this relation, we can interrupt an evaporation
sequence very close to the
BEC threshold, and record the instantaneous ion rate as well as
the corresponding TOF signal.  We typically chose the interruption
time for the evaporation ramp in advance, and then discarded the
runs for which the interruption time was further than $\pm$~60~ms
away from the threshold curve.

\begin{figure}
\begin{center}
\includegraphics[height=4cm]{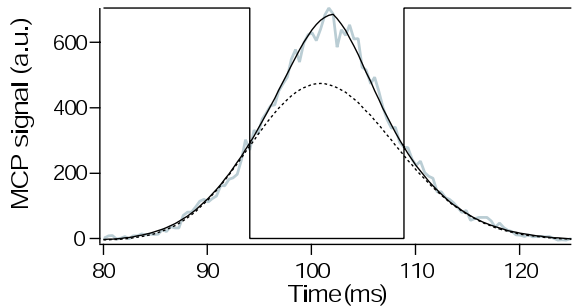}
\end{center}
\caption{
Time of flight signal corresponding to a cloud released from the
trap (at $t=0$) when its ion rate is on the threshold curve.
We fit the data with an excluded window indicated by the vertical
lines (width equal to the rms width of the TOF signal). A Gaussian
function (dotted line) does not describe the central part of the
data well, while the Bose function as defined in the text (solid
line), does, indicating that the cloud is close to threshold.}
\label{tofTsigma}
\end{figure}

We fit the associated TOF spectrum to determine the temperature
(Fig.~\ref{tofTsigma}). We use Eq.~(\ref{formuledensite}) together
with $\mu_{\rm c}$ given in Eq.~(\ref{mu}) for the initial atomic
density and then assume purely ballistic expansion of the cloud
after the switching off of the trap. We shall refer to this fit as
the Bose fit. The fits are weighted by an estimated uncertainty in
each point of the TOF curve.
To make this estimate, we chose a set of TOF spectra which
appeared to show no systematic deviation from their fits and used
them to estimate the amplitude of the noise. This noise varies as
the square root of the amplitude of the signal indicating that we
are limited by the shot noise of the atom detection. Our procedure
is
only an approximate indicator of the error bars. The chi square
per degree of freedom $\chi^{2}$ for the fits deduced in this way
ranges from 0.8 to 3. In addition, we first fit the TOF signal in
its entirety and then exclude a successively larger window of the
center of the spectrum out to the rms width of the spectrum. For
all of our runs, we observe a variation of less than 5\% and in
most cases less than 3\% of the fitted temperature as the excluded
window is increased.

In Fig.~\ref{tofTsigma}, we show an example of a typical TOF
spectrum and two fits to the data, using a Gaussian and the Bose
function described above. Both fits shown use only the wings of
the distribution. The ability of the Bose function to reproduce
the center of the distribution without including it in the fit,
unlike the Gaussian, confirms that the cloud is indeed close to
the condensation threshold.
In the following, we use the temperature given by the fit with an
excluded window of half the rms width of the TOF signal in order
to avoid the possibility of a small condensate component or other
high density effects distorting our analysis.


Before plotting the ion rate as a function of the critical
temperature, we must correct the observed temperature to account
for the hydrodynamic expansion of the cloud (see \cite{fabrice}
and references therein). This correction is done in the spirit of
Ref.~\cite{PGOS} which uses the Boltzmann equation approach to
take into account the effects of collisions during the expansion
of the cloud. The collision rate in Ref.~\cite{PGOS} is calculated
using a Gaussian density profile. We rather use the value
calculated for an ideal Bose gas \cite{Kavoulakis}, which we have
adapted to take interactions into account.
This correction depends on the scattering length but the effect on
the final value of $a$ is only of order 0.3~nm for $a$ ranging
from 10 to 14~nm. We therefore simply assume $a=12$~nm for this
correction in the following. In fact, due to the additional
anisotropy of the expanding cloud in the horizontal (detector)
plane, the fitting function should be modified; but a simulation
of this effect shows that the correction to the temperature is
less than 0.1 \%.

Finally, we correct the detected ion rate $\Phi_{{\rm c
},\,{\rm{det}}}$ to account for the detection efficiency $\alpha$
such that $\Phi_{\rm c}=\Phi_{{\rm c},\,{\rm{det}}}/\alpha$. It
should be noted that the rate constants were obtained by ion rate
measurements \cite{PRLHe}. This means that they were also
corrected: $\beta=\beta_{\rm{det}}/\alpha'$ and
$L=L_{\rm{det}}/\alpha'$. Equation \ref{formuleflux} shows that,
as long as $\alpha=\alpha'$, the detection efficiency cancels out
and do not have any impact on the determination of $a$. We have
checked experimentally that $\alpha=\alpha'$. To allow comparison
with figures in earlier publications, the ion rate and the
theoretical curves shown in all the figures have been corrected
using the same $\alpha$ as earlier, namely $\alpha=0.42$
\cite{PRLHe,Seidelin}.

\begin{figure}
\begin{center}
\includegraphics[height=5cm]{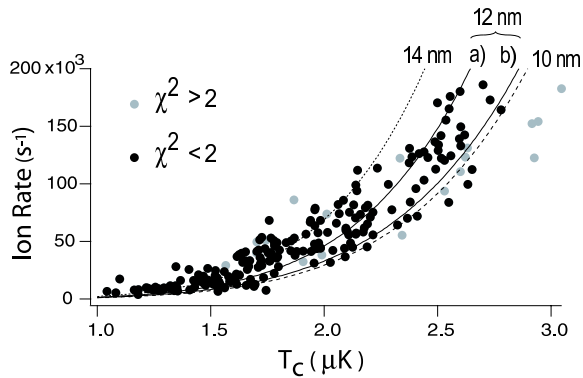}
\end{center}
\caption{Ion rate versus critical temperature. The points
correspond to the results of 280 runs for which the ion rate was
deemed sufficiently close to the condensation threshold. Gray
indicates runs for which $\chi^{2}$ in the TOF fits was above 2.
The dashed line is the theoretical estimate for $a=10$~nm, the
dotted line for $a=14$~nm (both including interaction corrections
of Eq.~\ref{formuleflux}). The two solid lines correspond to
$a=12$ nm, a) with interactions and b) without interactions, and
illustrate the size of their effect.}
\label{allonly}
\end{figure}

The results of this analysis are plotted in Fig.~\ref{allonly}.
Curves corresponding to the expected variation for three values of
the scattering length are also shown. We see that a large fraction
of the data falls between $a=10$ and 14~nm. The points at the
highest temperatures however, show a tendency to fall near the
theoretical curve for $a=10$~nm, while those at lower temperatures
fall near $a=14$~nm. To analyze this tendency further we examine
the TOF fits more closely using the $\chi^{2}$ value as an
indicator of the confidence level of each measurement. A large
$\chi^{2}$ could mean that the Bose function with $\mu$ imposed to
$\mu_{\rm c}$ is not the right fit function, and therefore that
the cloud is not sufficiently close to $T_{\rm c}$.
As shown in Fig.~\ref{allonly}, outliers tend to be correlated
with a large $\chi^2$.
Note however, that the remaining scatter in the data is too large
to be accounted for by our {\it a priori} estimates of the
uncertainties in our ion rate or temperature measurements, and we
presume that it is due to fluctuations in the determination of the
BEC threshold.

To determine the scattering length, we fit the black points in
Fig.~\ref{allonly} with $a$ as a free parameter and using $\beta$
and $L$ parameterized by $a$ as in Ref.~\cite{PRLHe}. The fit
gives (all points are given equal weight) $a=11.3$~nm. Our chief
estimated uncertainty stems from the fact that our data show a
systematic tendency to fall above the best fit at low temperature
and below it at high temperature. To estimate this uncertainty, we
fit the data (including gray points) separately for $T_{\rm c}$
below and above $2~\mu$K. We find $a=13.8$~nm for the low
temperature data and $a=10.4$~nm for the high temperature data.
The uncertainties in the measurements of $\beta$ and $L$ also
contribute to the uncertainty in Eq.~(\ref{formuleflux}) used for
fitting. In fact, the uncertainties in $\beta$ and $L$ are highly
correlated \cite{PRLHe} and their contribution to the uncertainty
is less than $0.5$~nm.

The error bars are obtained by summing quadratically the sources
of uncertainties. Our final result for the scattering length is
thus $a=11.3^{+2.5}_{-1.0}$~nm. This result may be compared with
the calculation of Ref.~\cite{Leo:2002}. This work leads to
$a=8$~nm using the potential curves of Ref.~\cite{Stark}. From
Ref.~\cite{Leo:2002} one also finds that a 0.5\% shift of the
repulsive part of that potential, would bring the theoretical
value into agreement with our result. This 0.5\% shift correspond
to the estimated uncertainty in the potential of Ref.~\cite{Stark}.
Another theoretical treatment \cite{Gadea} gives a scattering
length between 8 and 12 nm, also consistent with our results.

Our result also allows one to give values for the 2 and 3-body
ionization rate constants. The error-bars of Ref. \cite{PRLHe} are
modified to take into account the uncertainty of $a$. The
uncertainty in the ion detection efficiency also contributes to
the uncertainty in the rate constants. As discussed in Ref.
\cite{PRLHe}, we will assume $\alpha=0.42$ to get the central
values of the rate constants. We will include a one-sided
contribution to the error-bars to account for the possibility,
also discussed in \cite{PRLHe}, that $\alpha$ could be a factor of
2 smaller. We finally get $\beta=0.9^{+1.7}_{-0.8}\times
10^{-14}~{\rm{cm^3/s}}$ and $L=2.5^{+4.5}_{-1.7}\times 10^{-27}
~{\rm{cm^6/s}}$. The rate constants are in good agreement with
theoretical predictions \cite{Leo:2002,betaL}.

As shown in Fig.~\ref{allonly}, curves a) and b), our value of $a$
is significantly shifted by the non-ideal gas corrections of
Eq.~(\ref{formuleflux}). Thus, when an independent measurement of
the scattering length becomes available, our results can be used
as a test of these corrections \cite{Leducp}. Note however, that
corrections to the critical
temperature beyond mean-field theory \cite{beyond_m_f} are small
when one parametrizes the critical point in terms of an average
density \cite{Arnold2}. But an examination of the critical density
measured in a local way, by imaging the ions from a cloud for
example, is sensitive to critical fluctuation phenomena which go
beyond mean field theory similar to the homogenous case
\cite{beyond_m_f}. Thus, refinements of the ionization
measurements described here promise to continue to provide new
tests of BEC physics.

This work is supported by the EU under grants IST-2001-38863 and
HPRN-CT-2000-00125, and by the INTAS project 01-0855. SS
acknowledges support from the Danish Research Agency and JVG from
the Portuguese Foundation for Science and Technology (FCT).

\end{document}